\documentstyle[12pt,epsfig]{article}                 
\def\f {\cite{fad}~} 
\def\m {\cite{manu1}}
\def\mm {\cite{manu2}} 
\def\mmm {\cite{manu1,manu2}} 
\def\ms {\cite{manu3}} 
\def\be {\begin{eqnarray}} 
\def\ee {\end{eqnarray}} 
 
\def\n {{\hat{\bf{n}}}} 
\def\x {\hat{{\bf {\xi}}}} 
\def\no {\noindent} 

\begin{document}
\begin{flushright}
SNB/ 98-07-06  \\
\end{flushright}
\vskip1.5cm

\begin{center}
{\Large {\bf ABELIANIZATION OF LOW ENERGY SU(2) EFFECTIVE ACTION}} \\[15mm]

{\large Manu Mathur\footnote{Email:manu@boson.bose.res.in}}  \\[5mm]
{\em S. N. Bose National Centre for Basic Sciences, JD Block, \\
Sector III, Salt Lake City, Calcutta 700091, India. } \\[20mm]
\end{center}

\begin{abstract}
Recently Faddeev and Niemi proposed a low energy effective action 
for pure SU(2) Yang Mills theory in 4 dimension to describe its 
long distance physics. 
The effective action is O(3) $\sigma$ model with a  mass parameter,  
a dimensionless coupling constant e and a topological term.  In this work, 
choosing a new set of variables, we relate this $\sigma$ model 
to a U(1) gauge theory with electric and magnetic charges of 
charge  e and $4 \pi e^{-1}$ respectively. In the new formulation 
the connection of the mass parameter with the monopole condensate  
is discussed. This theory after lattice regularisation is the 
compact U(1) gauge theory coupled to electric charges. 

\end{abstract}

\newpage 

\begin{center}
1. \bf INTRODUCTION \\
\end{center}
\bigskip

Perhaps one of the most outstanding problems in physics is to 
understand the mechanism of color confinement in non-abelian 
gauge theories.  The idea of color confinement 
through dual Meissner effect was proposed by t' Hooft and Mandelstam 
\cite{thooft} more than twenty years back and is well accepted by now. 
This conjecture has been tested in the simpler toy model case of 
compact lattice quantum electrodynamics (CLQED). 
However, in the non-abelian gauge theories 
we still lack a general consensus on even the type of dynamical 
variables appropriate to describe its low energy physics. This can be contrasted 
with CLQED where  in the Villain form of the action the variables are 
the  photons and the magnetic monopoles  described 
by the vector field $A_{\mu}$ and an anti-symmetric tensor 
field $H_{\mu\nu}$ respectively. In terms of these variables the 
condensation mechanism of magnetic monopoles has been well understood 
by both analytical as well as Monte Carlo simulations \cite{pol}. 
In the strong coupling region the monopole condensation leads to 
dual Meissner effect confining all the electric charges. 
Motivated by this simple observation in the abelian case, in 
\m  we mapped the SU(2) gauge theory coupled to adjoint Higgs to 
$U(1)$ gauge theory by a change of variables in the partition 
function. The final dynamical variables were SU(2) {\it  gauge 
invariant} and included a ``photon" field topologically coupled 
to magnetic monopoles (like in CLQED) and minimally to 
electrically charged matter. Later, these results were 
generalised to SU(N) gauge  theories \mm. 

Recently, Faddeev and Niemi \f  proposed a reformulation of pure 
SU(2) Yang Mills theory in terms of new variables suitable for 
its low energy content. In terms of these variables 
the long distance physics is described 
by an  effective action which is O(3) $\sigma$ model with a 
a topological term and a mass parameter which appeared invoking 
the renormalisation group arguments. It was argued in \f   that 
this effective model describing infrared SU(2) physics is {\it unique} 
in the sense that it contains all the infrared relevant and marginal 
local Lorentz invariant operators of the $\sigma$ model field which are 
atmost quadratic in time derivative. However,  in order to understand 
the picture of color confinement  in this theory, it is desirable 
to explore  its connection  with the magnetic monopoles. 
In particular, an important issue is whether 
the mass parameter introduced  in \f   related to the monopole 
condensate. In this work we address these questions and show 
that the low energy  action of Faddeev and Niemi for SU(2) 
Yang Mills theory can be further  mapped into a U(1) gauge 
theory\footnote{This U(1) gauge group  has nothing to do with the initial
SU(2) gauge group.} with magnetic monopoles and electrically charged particles. 
This also establishes a connection  between \cite{fad} and the 
work done in \mmm. Moreover, using duality transformations,  we 
give a qualitative argument regarding the connection between the 
mass parameter and monopole condensate.  The plan of the paper is as follow: 
In the first part of the paper we will briefly discuss the kinematical 
issues related to the new set of variables which map $O(3)$ $\sigma$ 
model of \f   to the abelian gauge model. A more detailed discussion on the kinematical 
properties of these variables can be found in \mmm. In the second 
part we will study the dynamical content of the reformulated U(1) gauge 
theory  and argue that the new variables should be more appropriate 
to address the issues discussed above. 
 
The effective action describing pure SU(2) Yang Mills theory in \f    is: 

\begin{eqnarray}
S(\n) = \int_{x^4} \left[  m^2 (\partial_\mu \hat{\bf{n}}(x))^{2}   + 
{1 \over e^2} ( \hat{\bf{n}}(x) , \partial_{\mu} \hat{\bf{n}} (x) \times 
\partial_{\nu} \hat{\bf{n}} (x))^2 \right] . 
\label{ea}
\end{eqnarray}
 
\noindent In (\ref{ea}) $\hat{\bf{n}}(x)$ is a 3-dimensional unit 
vector in the internal SO(3) space, e is dimensionless coupling 
constant and m is the mass scale introduced 
invoking the renormalisation group arguments. While the 
details of (\ref{ea}) can be found in \cite{fad}, an important point for our purpose 
is the global invariance of (\ref{ea}) under the SO(3) transformations: 

\begin{eqnarray}
\hat{\bf{n}} (x) \rightarrow  \hat{\bf{n}} (x) + \vec{\lambda} \times  
\hat{\bf{n}} (x) . 
\label{o3} 
\end{eqnarray} 

\noindent 
\no The invariance of (\ref{ea}) under the transformation (\ref{o3}) corresponds 
to the global  SU(2) invariance of the initial standard Yang Mills action in terms of 
the gluon fields \cite{fad}. 
The low energy partition function is  

\begin{eqnarray} 
Z = \int {\cal{D}}(\hat{\bf{n}}) exp - S(\hat{\bf{n}}) . 
\label{parfun} 
\end{eqnarray}

\no In (\ref{parfun}) ${\cal{D}}(\hat{\bf{n}})$ is the SO(3) Haar measure 
and can be written as $sin\theta d\theta d \psi$, where $\theta(x)$  
and $\psi(x)$ are the polar and azimuthal angles of $\hat{\bf{n}}$ 
in the internal space.  We can also characterise $ \n$  in terms of its 
Eucledian co-ordinates in an internal frame rigidly attached in 
the space time. We call it space 
fixed frame (SFF)  and denote its orthonormal unit basis vectors by 
$\hat{e}^{a}$ (a =1, 2, 3). Generally, the dynamics is described 
in the SFF.  Given $\n (x)$,  we now construct a local body fixed 
frame (BFF) characterised by an orthonormal 
set of unit vectors ${\x}^{a}(x)$, (a=1, 2, 3) with the    
identification ${\x}^{3}(x)  \equiv \n (x)$. The other 
two basis vectors $\x^{\pm}(x) \equiv 2^{-1} (\x^{1}(x) \pm i \x^{2}(x))$ 
are arbitrary upto U(1) {\it local} rotations around $\x^{3}(x)$ axis. 
The BFF and SFF are related by a SO(3) matrix O(x): 

\begin{eqnarray}
\x^{a}(x) = O(x)^{a}_{b}~  \hat{e}^{b}  . 
\label{bs} 
\end{eqnarray} 
   
Any vector $\vec{v}(x)$ can also be expanded in the BFF: $\vec{v} \equiv 
v^{3} \x^{3} + v^{+} \x^{-} + v^{-} \x^{+}$\footnote{In what follows, 
the components $v^{\pm}$ will be called the chiral components of 
the vector ${\vec{v}}$.}. The  inbuilt local U(1) gauge invariance  
of the BFF in turn induces a local U(1) 
rotations on the BFF components of the vector 
\footnote{Note that the U(1) local transformations are   
change of basis in the internal space. Therefore 
all vectors are untouched and only their chiral BFF 
components undergo  induced rotations.} ${\vec{v}}$:

\begin{eqnarray} 
\x^{\pm}(x) \rightarrow exp(\pm i \alpha(x))\x^{\pm}(x) ~~ => ~~  
{v}^{\pm}(x) \rightarrow exp(\pm i \alpha(x)){v}^{\pm}(x).
\label{ab}
\end{eqnarray}

\no In (\ref{ab}), the local transformations being rotation, $\alpha(x)$ is 
compact. This construction of the BFF and the description 
of the dynamics (\ref{parfun}) in the BFF can be motivated by the two simple 
observations: 

\begin{enumerate}

\item The dynamics (\ref{parfun}) when described in the BFF will have 
an inbuilt local U(1) gauge invariance.  In the sequel, we will show 
that this reformulated dynamics and the associated local U(1) gauge invariance is 
that of ``photons" minimally (topologically) coupled to electric 
(magnetic) charged matter fields. 

\item The components of all vectors in the BFF are explicitly 
invariant under (\ref{o3}). This SO(3) invariance also implies 
invariance under the initial SU(2) global transformations. 
\end{enumerate} 

In order to describe the dynamics of (\ref{parfun}) in the BFF, we 
now make a new change of variables from the description  
$\n(x)   (\equiv \x^{3}(x))$ to the  set of vector fields 
$\vec{\omega}_{\mu}$ defined by: 

\begin{eqnarray}
\partial_{\mu} \hat{\xi}^1 \equiv \omega_{\mu}^{2} \hat{\xi}^3 - 
\omega_{\mu}^{3} \hat{\xi}^2, ~~~~ \partial_{\mu} \hat{\xi}^2 
\equiv \omega_{\mu}^{3} \hat{\xi}^1 - \omega_{\mu}^{1} \hat{\xi}^3, ~~~~
\partial_{\mu} \hat{\xi}^3 \equiv \omega_{\mu}^{1} \hat{\xi}^2 - 
\omega_{\mu}^{2} \hat{\xi}^1. 
\label{angvel}
\end{eqnarray}

\no The eqns. (\ref{angvel}) are nothing but the orthonormality of 
the BFF and can be re-written 
in compact form $D_{\mu}(\vec{\omega}) \hat{\xi}(x) \equiv 0$ where 
the  ``covariant derivative" $D_{\mu}^{ac}(\vec{\omega}) 
\equiv \delta^{ac}\partial_{\mu} - \epsilon^{abc}\omega_{\mu}^{b}$ 
is defined with respect to the new variables 
$\vec{\omega}_{\mu}$. Infact,  these  new variables 
have a simple geometrical interpretation of the ``angular velocities" 
of the  BFF with respect to the SFF \m.  
The $U(1)$ gauge transformations (\ref{ab}) on the BFF components of  
$\vec{\omega}(x)$ are given by: 

\begin{eqnarray}  
\omega_{\mu}^3(x)  \rightarrow  \omega_{\mu}^3(x) + \partial_{\mu} 
\alpha(x), ~~~~~~    \omega_{\mu}^{\pm}(x)  \rightarrow   
exp(\pm i \alpha(x)) \omega_{\mu}^{\pm}(x). 
\label{angvelgta} 
\end{eqnarray} 

\no From (\ref{angvelgta}) we see that $\omega^{3}_{\mu}$ transforms 
like ``photon" and the orthogonal chiral components of the angular 
velocities transform like electrically charged matter fields. 
Therefore, we will denote\footnote{The coupling 
constant e in the definition of $A_{\mu}$ has been introduced for later 
convenience.} $\omega^{3}_{\mu}$ by $e A_{\mu}$. 
We can now define the U(1) co-variant derivative of 
$\omega^{\pm}_{\mu}$: 

\begin{eqnarray} 
D_{\mu}(A) \omega^{\pm}_{\mu} \equiv \left(\partial_{\mu} \pm i e 
A_{\mu}\right) \omega^{\pm}_{\mu}
\label{coder} 
\end{eqnarray} 

\no We also have the original SO(3) global transformations: 

\begin{eqnarray} 
\vec{\omega}_{\mu}(x) \rightarrow \vec{\lambda} \times \vec{\omega}_{\mu}(x)
\label{o32} 
\end{eqnarray} 

 The final dynamical fields in the reformulated model  
will be the angular velocities $\vec{\omega}_{\mu}$. The eqn. 
(\ref{o32}) implies that in the BFF we will have a formulation 
in terms of explicitly SU(2) color neutral fields. 
In (\ref{parfun}), we have replaced 2 compact degrees of freedom of the 
initial variable $\hat{n}$ by a set of 12 non-compact angular velocities. 
Therefore, not all of them are independent. 
The corresponding constraints are  easily obtained by defining the  
``field strength tensor"  $ \vec{{F}}_{\mu \nu}\left(\vec{\omega}(x)\right) \equiv 
\partial_\mu \vec{\omega}_\nu - \partial_\nu \vec{\omega}_\mu + \vec{\omega}_\mu 
\times \vec{\omega}_\nu$ and  considering the commutator:

\begin{eqnarray}  
[D_{\mu}(\omega),D_{\nu}(\omega)] \hat{\xi}^{a} = 0 => \vec{F}_{\mu\nu}
\left(\vec{\omega}(x)\right) = \vec{L} cos \theta(x) 
\left(\partial_{\mu}\partial_{\nu} - \partial_{\nu}\partial_{\mu}\right) 
\psi(x) 
\label{con} 
\end{eqnarray} 

In (\ref{con})  $\vec{L} \equiv (sin\theta,~ 0,~ - cos\theta)$.   
The space time points where the azimuthal angle $\psi(x)$ is single valued 
the right hand side of (\ref{con}) vanishes and the solutions are pure gauges.  
From (\ref{bs}) and (\ref{angvel}) we find that $\omega_{\mu}(x) = O(x) 
\partial_{\mu} O^{-1}(x)$. This orthogonal matrix O(x) defined in 
(\ref{bs}) can be characterised by the 3 Euler angles $(\alpha,\theta,\psi)$. 
Thus we recover the original 2 angular degrees of freedom of $\n$ 
along with the U(1) gauge angle. We now come to the hidden topological 
degrees of freedom in (\ref{parfun}). These degrees of freedom are at 
the space time points $[x_{0}]$ where the R.H.S of (\ref{con}) is 
non-vanishing. These points can be characterised by the topological index: 

\begin{eqnarray} 
N = {1 \over 2\pi} \int_{{\cal C}=\partial \Sigma_{x_{0}}} d x_{\mu} 
\partial_{\mu} \psi(x) = 2 \pi {\cal Z}  
~~~~~~~~{\cal Z} \in Integers. 
\label{top} 
\end{eqnarray} 

\no In (\ref{top}) $\Sigma_{x_{0}}$ is an infinitesimal area enclosing 
the point $x_{0}$ and ${\cal C}$ is its boundary. 
The necessary condition for the R.H.S of (\ref{top}) to be non-zero is  
$\theta(x_{0}) = 0/ \pi$. Equivalently the unit vector $\hat{\bf{n}}$ 
is rotated along either the +ve or -ve polar axis in the internal space\footnote{This 
implies $\vec{L}$ in (\ref{con}) is effectively $(0, 0, \pm 1)$}. Therefore,  
$[x_{0}]$ form a one dimensional string. We will see that these strings 
correspond to monopoles attached with Dirac strings carrying  $4 \pi  e^{-1} 
{\cal Z}$ unit of  magnetic  flux towards the monopoles . To study 
this aspect further we re-write the constraints 
(\ref{con}) in a  form which will be more convenient at later stage: 

\begin{eqnarray}  
F^{\pm}(\omega)_{\mu\nu}  = D_{\mu}(A) \omega^{\pm}_{\nu} - D_{\nu}(A) 
\omega^{\pm}_{\mu} = 0 ~~~~~~~~~~~ \nonumber \\ 
\Big(\partial_{\mu} A_{\nu} - \partial_{\nu} A_{\mu}  - {cos\theta \over e} 
\left(\partial_{\mu}\partial_{\nu} 
- \partial_{\nu}\partial_{\mu}\right)\psi\Big)
 = {2 i \over e} \left(\omega^{+}_{\mu} \omega^{-}_{\nu} - h.c\right) . 
\label{conc} 
\end{eqnarray} 

We now consider the dynamical issues related to (\ref{parfun}). The 
action (\ref{ea}) in terms of angular velocities is:  

\begin{eqnarray} 
S(\n) = 4 m^{2} \omega_{\mu}^{+} \omega_{\mu}^{-} + \Big({2 i \over e}\Big)^{2} 
\Big(\omega^{+}_{\mu} \omega^{-}_{\mu} - h.c\Big)^{2}. 
\label{ean} 
\end{eqnarray} 

\no We notice that the topological term in (\ref{ean}) is just the square of 
the left hand side of the constraint (\ref{conc}). Therefore,  we can write:

\begin{eqnarray} 
S(\n) =  S(A_{\mu}, \omega^{\pm}_{\mu}, \psi) = \Big[\partial_{\mu} 
A_{\nu} - \partial_{\nu} A_{\mu}
- {\cal F}^{np}_{\mu\nu}\Big]^{2} + 4 m^{2} \omega_{\mu}^{+} \omega_{\mu}^{-} 
\label{eann} 
\end{eqnarray} 

\no In (\ref{eann}) ${\cal F}^{np}_{\mu\nu} \equiv  e^{-1} cos\theta  
\left(\partial_{\mu}\partial_{\nu}- \partial_{\nu}\partial_{\mu}\right)\psi$ 
and the superscript (np) is to emphasize that it is a non-perturbative term.
We can now define the abelian field strength tensor: 

\begin{eqnarray} 
F_{\mu\nu}(A) \equiv \partial_{\mu} A_{\nu} - \partial_{\nu} A_{\mu} - 
{\cal F}_{\mu\nu}^{np} 
\label{u1fst} 
\end{eqnarray} 

\no Along the one dimensional space time point 
where $\theta$ = 0 or $\pi$ and $\psi$ is multivalued, we 
get a  singular magnetic flux from 
${\cal F}_{\mu\nu}^{np}$ in (\ref{u1fst}). On these 
singular magnetic strings the value of $\theta$ can flip from 
0 to $\pi$ and thus changing the direction of the magnetic 
flux. 
As the polar angle $\theta$ of $\n$ is not invariant under  
the transformations (\ref{o3}), the $\theta$ = 0 and 
$\theta$ = $\pi$ parts of the strings are unphysical. Their 
locations can be changed by the transformation (\ref{o3}). 
However, the discrete space time points on a particular string 
where $\theta$ flips between 0 and $\pi$ are left invariant 
under global SO(3) transformations and are the sources of 
magnetic flux.    
More explicitly,  defining the topological magnetic 
current $K_{\mu} \equiv \partial_{\nu}\tilde{{\cal F}}^{np}_{\mu\nu}$ 
($\tilde{F}_{\mu\nu} \equiv 
2^{-1}\epsilon_{\mu\nu\rho\sigma}F_{\rho\sigma}$). The Bianchi 
identities are:  

\begin{eqnarray} 
\partial_{\nu}\tilde{F}_{\mu\nu}  \equiv K_{\mu}.  
\nonumber 
\end{eqnarray} 

Thus the first term in the action (\ref{eann}) has the simple physical 
interpretation of photons interacting with SO(3) invariant magnetic 
monopoles of charge $(4 {\cal Z} \pi e^{-1})$ attached with two 
unphysical Dirac strings charecterised by 
$\theta = 0$ and $\theta =\pi$ respectively. Each of them carry 
$2 \pi {\cal Z} e^{-1}$ unit of the magnetic flux towards the monopole. 
It is interesting to note that the origin of these topological magnetic 
monopoles is the topological term in (\ref{ea}). 

The partition function (\ref{parfun}) in terms of the U(1) gauge and 
charged matter is: 

\begin{eqnarray} 
Z  =  \int dA_{\mu} d\omega^{\pm}_{\mu} d \theta d \psi ~~\delta(C^{3}(x))
 ~\delta(C^{+}(x)) ~\delta(C^{-}(x))  
 ~exp - S \left(A_{\mu}, \omega^{\pm}_{\mu}, \psi\right) ~~ \nonumber \\  
S \left(A_{\mu}, \omega^{\pm}_{\mu}, \psi\right)   =   
\int_{x^{4}} \left[ 4 m^{2} \omega_{\mu}^{+} \omega_{\mu}^{-} + 
\Big(\partial_{\mu} A_{\nu}-\partial_{\nu} A_{\mu} - 
{\cal F}^{np}_{\mu\nu}\Big)^{2}\right] 
 ~~ \label{parfunn}
\end{eqnarray}

\no In (\ref{parfunn}) we have used $C^{\pm}$ and $C^{3}$ to denote the 
three constraints in (\ref{conc}) respectively.  We once again mention the two 
important properties of  (\ref{parfunn}): a) It is manifestly  
invariant under local U(1) gauge transformations,    
b) All the fields appearing in (\ref{parfunn}) are 
SU(2) color neutral. Both these properties were expected 
and were emphasized in the begining.  

To handle the singular fluxes in ${\cal F}^{np}_{\mu\nu}$, the 
most convenient regularization 
is on the lattice. On lattice the minimal length curve is the boundary 
of a plaquette. Therefore the multivalued field $\psi$ and their 
derivatives are well defined. In other words the singular integer 
valued flux (in the units of $4 \pi e^{-1}$) along the strings gets 
spread  over a plaquette.
This integer flux can be characterised by an anti-symmetric tensor 
field $H_{\mu\nu}$ \cite{manu3} and the partition function can be written 
as: 

\begin{eqnarray}
Z = \sum_{H_{\mu\nu} = -\infty}^{+\infty} \int dA_{\mu} 
d\omega_{\mu}^{\pm} \delta(C^{\pm})\delta(C^{3}) 
exp - S(A_{\mu}, H_{\mu\nu}) ~~~~  \label{finalm} \\  
S(A_{\mu}, H_{\mu,\nu}) = \int_{x^4} \Big[\Big(\partial_{\mu} A_{\nu} 
- \partial_{\nu} A_{\mu} 
- {4 \pi \over e} H_{\mu\nu}\Big)^{2} + 4 m^{2} \omega_{\mu}^{+} 
\omega_{\mu}^{-}\Big]  
\nonumber 
\end{eqnarray} 

\no In (\ref{finalm}) the integer valued anti-symmetric field 
reflects the multi-valued nature of the field $\psi$. 

All the fields in (\ref{finalm}) are SU(2) color neutral 
by construction. However, now we have the new U(1) symmetry 
whose charges must be  unobservable. They should be confined 
via dual Meissner effect. At this 
stage it is interesting to compare (\ref{finalm}) with the 
results of \cite{manu3} where a manifestly Lorentz co-variant and local 
quantum field theory for magnetic monopoles and electric 
charges was proposed by exploiting duality transformations. 
The starting theory was dual abelian Higgs model with a
complex scalar fields $(\phi_{m}(x), 
\phi^{*}_{m}(x))$ carrying magnetic charge (g) and coupled minimally 
to the dual vector potential\footnote{The subscript 
m on the complex scalar field is the emphasize that these fields 
carry the magnetic charges. The electric field in terms of the 
dual vector potential is given by $E_{i} \equiv \epsilon_{ijk} 
\partial_{j} \tilde{A}_{k}$} 
$\tilde{A}_{\mu}(x)$ in the presence of a potential $V(\phi_{m}
\phi^{*}_{m})$. The action is: 

\begin{eqnarray} 
S(\phi_{m}, \phi^{*}_{m}, \tilde{A}) 
= D_{\mu}(\tilde{A})\phi_{m} D_{\mu}(\tilde{A})\phi_{m}^{*} + {1 \over 4} 
\Big(\partial_{\mu}\tilde{A}_{\nu} - 
\partial_{\nu}\tilde{A}_{\mu}\Big)^{2} + \Big(\phi_{m} \phi^{*}_{m} 
- m^{2}\Big)^{2} .  
\label{adhm} 
\end{eqnarray} 

\no The co-variant derivatives in (\ref{adhm}) are defined in the 
standard way. Decomposing the complex scalar field into its radial 
and angular parts $\phi_{m} \equiv \rho(x) exp i \psi(x)$ and then 
performing a duality transformations, we found that the partition 
function of (\ref{adhm}) could be re-written as \ms : 

\begin{eqnarray} 
Z = \sum_{H_{\mu\nu}=-\infty}^{+\infty} \int \rho d\rho d A_{\mu} 
exp - S  \hspace{5cm}  \label{dhmdual} \\  
S = \int_{x^{4}} \Big[{1 \over 4}\Big(\partial_{\mu}{A}_{\nu} -
\partial_{\nu}{A}_{\mu} - g H_{\mu\nu}\Big)^{2} 
+ { 1 \over 4 {\rho}^{2}}  \Big(\partial_{\mu}\tilde{H}_{\mu\nu}\Big)^{2} 
+ \Big(\partial_{\mu}\rho\Big)^{2} + \Big({\rho}^2 - m^{2}\Big)^{2}\Big] .   
\nonumber  
\end{eqnarray} 

\no Note that the role of the multivalued field $\psi$, 
representing the phase of complex Higgs $\phi(x)$ in the abelian 
case is similar to its role in SO(3) $\sigma$ model case where 
it represented the azimuthal angle of $\n$. 
In both abelian and non-abelian cases 
its dual field is the antisymmetric tensor field representing 
the magnetic monopoles. This correspondence is also valid for 
SU(N) gauge theories \mm.  

We expect that the integration of charged matter field 
$\omega_{\mu}^{\pm}$ in (\ref{finalm}) will generate 
a term proportional to $(4 m^{2})^{-1} (\partial_{\mu} 
\tilde{H}_{\mu\nu})^{2}$. This term is consistant with 
the local U(1) gauge invariance of the theory. If true, 
comparing it with (\ref{dhmdual}) at $\rho$ = m, 
this effective theory derived from the 
original Yang Mills action via (\ref{parfun}) 
will describe an explicit SU(2) color neutral  dual superconductor 
(\ref{adhm})  in dual language (i.e in terms of $A_{\mu}$ and $H_{\mu\nu}$).  
This important issue is currently under investigation and will be 
reported elsewhere. 
Finally it will be interesting to study the massive knotlike solitons 
\cite{fad2} in  terms of the new variables discussed in this paper. 

\end{document}